\documentclass{article}
\usepackage{color}
\usepackage{PRIMEarxiv}
\usepackage{subfig}
\usepackage[utf8]{inputenc} 
\usepackage[T1]{fontenc}    
\usepackage[hidelinks]{hyperref}       
\usepackage{url}            
\usepackage{booktabs}       
\usepackage{amsfonts}       
\usepackage{nicefrac}       
\usepackage{microtype}      
\usepackage{lipsum}
\usepackage{mwe}
\usepackage{subfloat}
\usepackage{fancyhdr}       
\usepackage{graphicx}       
\graphicspath{{media/}}     
\usepackage{subcaption}
\usepackage{caption}
\usepackage{mathptmx}
\usepackage[affil-it]{authblk}

\title{AmarDoctor: An AI-Driven, Multilingual, Voice-Interactive Digital Health Application for Primary Care Triage and Patient Management to Bridge the Digital Health Divide for Bengali Speakers

}

\author[1]{\href{https://orcid.org/0009-0003-3708-9029}{Nazmun Nahar}}
\author[ ]{\href{https://orcid.org/0009-0007-4805-295X}{Ritesh Harshad Ruparel}}
\author[1]{\href{https://orcid.org/0009-0006-6749-8570}{Shariar Kabir}}
\author[1,3]{Sumaiya Tasnia Khan}
\author[1]{\href{https://orcid.org/0000-0001-8867-1770}{Shyamasree Saha}}
\author[1, 4,*]{\href{https://orcid.org/0000-0001-9321-0469}{Mamunur Rashid}}

\affil[1]{\href{https://medaihealth.com/}{MedAi Bangladesh Limited}}
\affil[3]{Central Police Hospital, Dhaka, Bangladesh}
\affil[4]{University of Birmingham, UK}
\affil[*]{Corresponding author: mamun.rashid@medaihealth.com}

\newpage

\begin{document}
\maketitle

\begin{abstract}
This study presents AmarDoctor, a multilingual voice-interactive digital health app designed to provide comprehensive patient triage and AI-driven clinical decision support for Bengali speakers, a population largely underserved in access to digital healthcare. AmarDoctor adopts a data-driven approach to strengthen primary care delivery and enable personalized health management. While platforms such as AdaHealth, WebMD, Symptomate, and K-Health have become popular in recent years, they mainly serve European demographics and languages. AmarDoctor addresses this gap with a dual-interface system for both patients and healthcare providers, supporting three major Bengali dialects. At its core, the patient module uses an adaptive questioning algorithm to assess symptoms and guide users toward the appropriate specialist. To overcome digital literacy barriers, it integrates a voice-interactive AI assistant that navigates users through the app services. Complementing this, the clinician-facing interface incorporates AI-powered decision support that enhances workflow efficiency by generating structured provisional diagnoses and treatment recommendations. These outputs inform key services such as e-prescriptions, video consultations, and medical record management. To validate clinical accuracy, the system was evaluated against a gold-standard set of 185 clinical vignettes developed by experienced physicians. Effectiveness was further assessed by comparing AmarDoctor performance with five independent physicians using the same vignette set. Results showed AmarDoctor achieved a top-1 diagnostic precision of 81.08 percent (versus physicians average of 50.27 percent) and a top specialty recommendation precision of 91.35 percent (versus physicians average of 62.6 percent).

\end{abstract}


\section{INTRODUCTION}
In an era where technology is integrated into nearly every aspect of daily life, healthcare has undergone a significant shift toward digitization.  Digital health solutions, including AI-powered symptom assessment tools, virtual consultations (e.g., telehealth, telemedicine), and lifestyle and wellness guidance, are now at the forefront of this transformation, becoming increasingly prevalent among patients who turn to the internet as a primary resource for health-related concerns.  AI-powered self-triaging solutions such as WebMD, Symptomate, Ada, and Klinik Health have played a key role in this digital health revolution, empowering individuals to take greater control over their healthcare decisions \textcolor{blue}{\cite{Yang2023}, \cite{Zawati2024}}, \textcolor{blue}{\cite{gilbert2020accurate}}. Amidst the growing demand for virtual primary care, healthcare systems worldwide are embracing these AI-powered digital triaging applications to achieve more efficient patient management  \textcolor{blue}{\cite{baker2020comparison}}. These cutting-edge tools are redefining primary care delivery by alleviating workload pressures on medical staff in many developed countries (\textbf{Klinik Healthcare, May 2021}).

However, despite these advancements and their rapid adoption in developed countries, digital health solutions face significant challenges in scaling equitably across diverse contexts. While these technologies hold the potential to address gaps in healthcare access – it's been estimated that half of the world's population lacks access to essential healthcare services \textcolor{blue}{\cite{Marcin2016}}, \textcolor{blue}{\cite{Labrique2018}} – there's growing evidence indicating that they are, in some cases, widening existing inequalities \textcolor{blue}{\cite{Dorsey2016}}, \textcolor{blue}{\cite{ScottKruse}}, \textcolor{blue}{\cite{Veinot2018}}. This is particularly evident in low- and middle-income countries, where various factors hinder the effective implementation and adoption of digital health solutions. In this context, the case of Bengali speakers is particularly salient. Bengali, the world's seventh-largest ethnolinguistic group, with 270 million speakers across Bangladesh, India, and the Gulf region, remains one of the most underserved populations in digital healthcare.  Especially, Bangladesh, with its 180 million population, has one of the worst primary care systems among its contemporaries, with only 9.9 doctors for every 10,000 patients \textcolor{blue}{\cite{nuruzzaman2022informing}}. The situation is even more severe in rural areas, where there are only 1.1 doctors for the same number of patients \textcolor{blue}{\cite{joarder2018retaining}}.  Due to the absence of an established primary care system, trivial illnesses often escalate into health emergencies, pushing 6 million people into poverty due to out-of-pocket (OOP) emergency health expenditure. Despite high smartphone penetration (60 million) and good internet coverage, digital health solutions are not scaling well in Bangladesh. 

Several primary barriers seriously hinder the scaling of digital health solutions in LMIC countries like Bangladesh.  

\textbf{Poor Digital Literacy}
These include poor digital literacy among both patients and doctors  \textcolor{blue}{\cite{Toscos2019}}, and the fact that the majority of available digital health solutions primarily support English or European languages, failing to encompass the linguistic nuances around the medical vocabulary for many non-European languages. Furthermore, digital health services available via mobile or web applications require patients to type in a digital medium (e.g., smartphone or keyboard). There are several versions of Bengali digital keyboards (e.g., Unicode-based Bengali keyboards) and a mixture of phonetic-based approaches, each with different layouts. This lack of standardization makes typing Bengali in digital media unpopular among low socioeconomic groups and those with poor digital literacy \textcolor{blue}{\cite{8122676}}.

\textbf{Bias in Current Clinical Decision Support Systems}
Furthermore, AI-driven clinical decision support systems offer the potential to help physicians reduce consultation time and increase operational efficiency, which is a critical priority in many developing countries. However, the effectiveness of these platforms is often limited by their reliance on patients' ability to accurately report their medical concerns through a text-based interface. This presents a significant challenge for many non-Latin language speakers with low digital literacy, as their languages frequently do not translate seamlessly onto standard digital keyboards. This issue is further compounded by the fact that current recommendation systems are often trained on data predominantly from white European patients, which can significantly reduce their accuracy and effectiveness when applied to patients from non-European ethnolinguistic backgrounds  \textcolor{blue}{\cite{Kerasidou2021}, \textcolor{blue}{\cite{Fletcher2021}}, \cite{Obermeyer2019}}. Consequently, these oversights can disadvantage native language speakers who lack adequate English skills and prefer to use their native tongues when accessing digital health solutions, thereby increasing the risk of exacerbating existing health inequalities \textcolor{blue}{\cite{al2020implications}, \cite{Veinot2018}}.

The convergence of these technical challenges with prevalent poor literacy has significantly impeded the progress of healthcare digitization and the widespread adoption of digital technologies. 

\textbf{Lack of Domain Specific Foundation Models in Bengali}

Recent advancements in large language models (LLMs) are beginning to reshape the digital healthcare landscape, offering unprecedented capabilities in clinical support and patient engagement. However, the efficacy of these powerful models is not universal; their performance is highly contingent on the linguistic and clinical context for which they are trained \textcolor{blue}{\cite{Singhal2025}}. This creates a significant disparity for the world's 270 million Bengali speakers. Despite its global prevalence, Bengali remains a low-resource language in the digital domain, with a pronounced lack of medical-specific foundational models, which has hindered the ability to adapt cutting-edge AI for this population \textcolor{blue}{\cite{Bhowmik2025}}.

Consequently, while multi-modal LLMs are empowering millions worldwide with access to sophisticated medical advice and user-friendly conversational bots, Bengali speakers have largely been excluded from these advancements, deepening the digital divide in healthcare access

\begin{figure}[!ht]
  \centering
  \includegraphics[width=16.5cm]{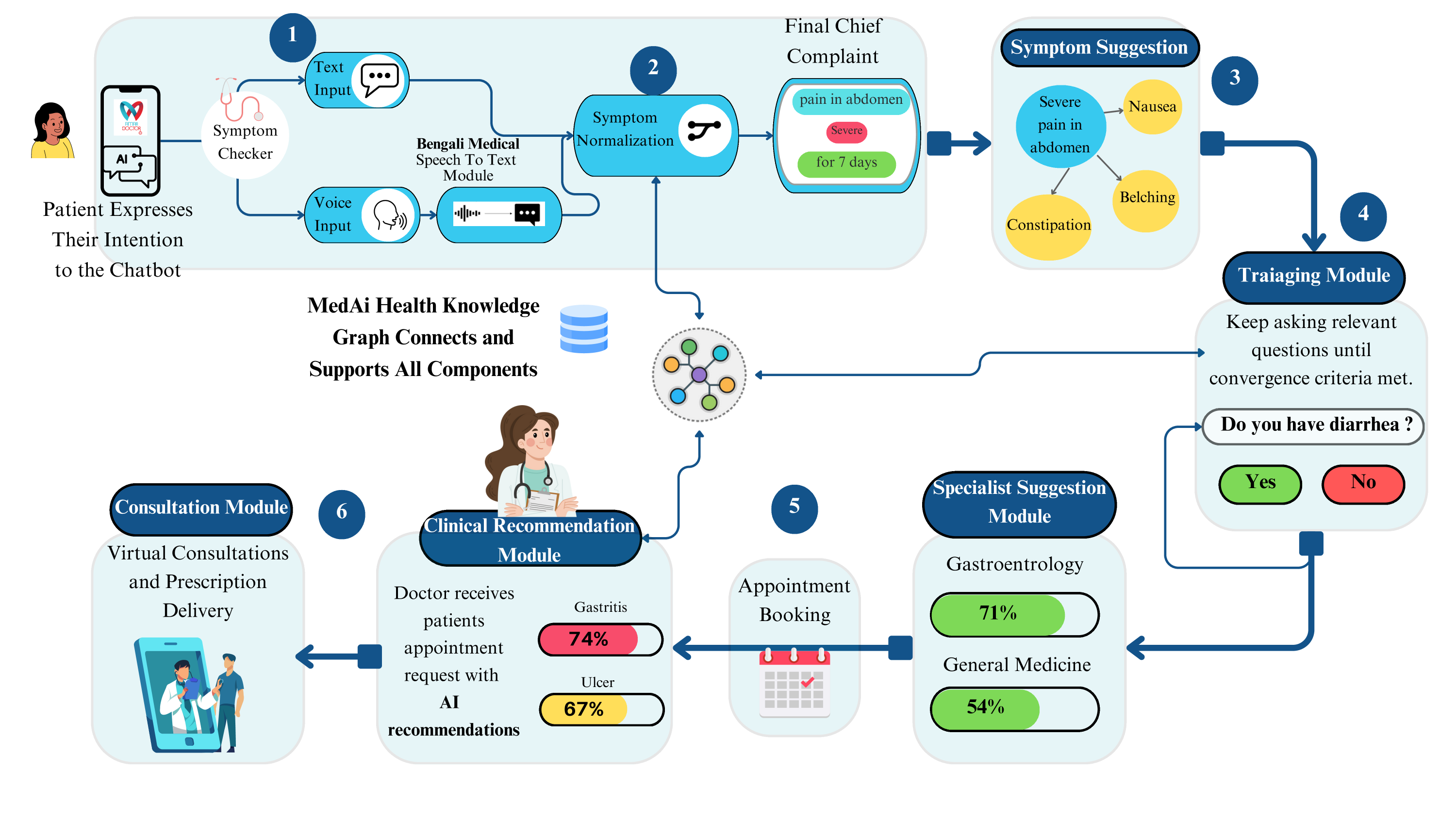}
  \caption{AmarDoctor's patient-centered medical triage and primary healthcare delivery. The workflow begins with a multilingual, voice-interactive chatbot that classifies a patient's intent to guide them to the appropriate service. For users reporting symptoms to book a doctor, the system initiates a six-stage process: (1) capturing chief complaints via text or voice input; (2) normalizing medical terminology using the MedAi Health Knowledge Graph; (3) suggesting related symptoms to the patient based on their initial input; (4) conducting an interactive diagnostic assessment with relevant medical questions leveraging the knowledge-graph; (5) generating specialist doctor suggestions for the patient and provisional clinical diagnoses for the physician ; and (6) facilitating virtual consultation booking, video consultation from a doctor and finally prescription delivery after the consultation.}
  \label{fig:figure 1}
\end{figure}

This paper presents AmarDoctor, a groundbreaking digital health platform developed to address the aforementioned challenges and deliver inclusive healthcare solutions for South Asian communities. AmarDoctor's core innovation lies in its application suite, which includes: (1) a multilingual, voice-enabled remote patient triaging system that facilitates accurate symptom collection in both Bengali and English, thereby overcoming language and literacy barriers, and (2) a dual-recommendation engine that provides two distinct outputs: guidance on the appropriate medical specialist for the patient, and an AI-driven provisional diagnosis for the clinician to enhance diagnostic efficiency. Underpinning these features is a comprehensive health knowledge graph, constructed using extensive patient data and refined by expert clinicians, ensuring broad coverage of South Asian symptom and disease vocabulary, including regional dialect variations. Our evaluation of the system's performance demonstrates promising accuracy in both provisional diagnoses and specialist referral recommendations. To the best of our knowledge, this work represents the first of its kind focusing on AI-driven clinical decision support tailored for Bengali speaking patients, setting a foundation for inclusive AI that aims to democratize healthcare access.

\section{METHODOLOGY}
\subsection{Construction of the Health Knowledge Graph }
The foundation of the AmarDoctor platform is a comprehensive health knowledge graph, meticulously constructed to model the intricate relationships among key medical entities such as patients' diseases, symptoms (phenotypes), prescribed drugs, and diagnostic tests. This graph was developed by integrating aggregated analytical insights from 1.4 million anonymized patient clinic visit records (encompassing symptoms, diagnoses, prescribed medications, and diagnostic tests) with proprietary patient data gathered from two pilot studies conducted in Bangladesh.

The resulting graph is composed of five distinct entities, or nodes, each with specific attributes: (i) person (e.g., patient), (ii) \textbf{279} unique diseases, (iii) \textbf{908} unique symptoms, (iv) \textbf{12,334} unique drugs, and (v) \textbf{11,051} procedures. The relationships between these entities are quantified by edge weights, which signify their relative importance. Initially, these weights were derived from the aggregated patient data analytics.

To address potential biases arising from the over-representation of European ancestry patient data in the initial dataset, a critical refinement process was undertaken. A panel of three expert clinicians specializing in general practice meticulously reviewed and adjusted the weights for disease-symptom, symptom-symptom, disease-drug, and disease-procedure relationships. This rigorous review, cross-referenced with published literature and clinical experience, ensured that the weights accurately reflect symptom prevalence and manifestation in diseases specific to the South Asian patient population. This meticulously refined health knowledge graph serves as the fundamental underpinning for several core components of the AmarDoctor application suite, including the symptom checker and the provisional diagnosis system.

\subsection{Voice-Interactive Multilingual Healthcare Assistant}
{AmarDoctor} introduced the first multilingual (Bengali and English) voice-interactive healthcare assistant  "Aisha" that allows patients to navigate various services within the app seamlessly. The chatbot deploys combination of proprietary medical speech recognition \cite{kabir2024automaticspeechrecognitionbiomedical} and customized Large Language Model (LLM) powered intent classification technique to identify patient's intentions (e.g. doctor booking, medical complaints). It offers an empathetic and professional voice-interactive and guides patients to the right service within the AmarDoctor app. This functionality allows users with poor digital literacy to easily adopt digital health solutions and timely access to primary healthcare. Currently, the health assistant covers 14 possible intents from finding nearest hospital, learn about medicine to book doctor for virtual medical consultation from home. 

\subsection{AI Guided Symptom Reporting}
\textbf{AmarDoctor} presents a novel multilingual symptom checker module designed to facilitate patient-reported medical concerns. By leveraging the comprehensive health knowledge graph (sec 2.1), the module guides patients through a structured series of medical questionnaires, effectively capturing their present symptoms. To significantly enhance accessibility for a diverse patient population, particularly those with poor digital literacy, it supports both English and Bengali (both standard and colloquial), including major dialects \textbf{Sylheti} and \textbf{Chittagonian}. A total of \textbf{908} unique symptoms were mapped to these dialects, ensuring that patients from various linguistic backgrounds can effectively communicate their health concerns. The user experience of the module is further bolstered by the integration of various input modalities, emphasizing a multi-modal approach: \textbf{(1)} a comprehensive curated symptom list (\textit{Figure }\ref{fig:fig3}), \textbf{(2)} common ailments such as fever, pain, cough, diarrhea, nausea, vomiting etc., \textbf{(3)} algorithm-suggested symptoms, \textbf{(4)} free text entry, and \textbf{(5)} a voice-interactive medium. This combination of voice, free text input, and visual suggestion of common symptoms is crucial to allow patients with poor digital literacy to access the benefits of digital health platforms. The voice-interactive module leverages our proprietary Bengali medical ASR module 
\cite{kabir2024automaticspeechrecognitionbiomedical}. This diverse range of options ensures that patients can effectively communicate their symptoms, regardless of their linguistic abilities or preferences. To mitigate speech-to-text transcription errors and handle variations in user input (e.g., spelling errors, use of synonyms, colloquial terms), a symptom mapping algorithm was implemented to align user-reported symptoms with our existing symptom ontology, using a semantic similarity-based searching strategy to match user-specified symptoms to the nearest symptom in our vocabulary. If a user cannot input any symptom correctly, the doctor has the option to add or update symptoms based on their conversation with the patient during the virtual consultation.
\begin{figure}[!ht]
  \centering
  \includegraphics[width=15cm]{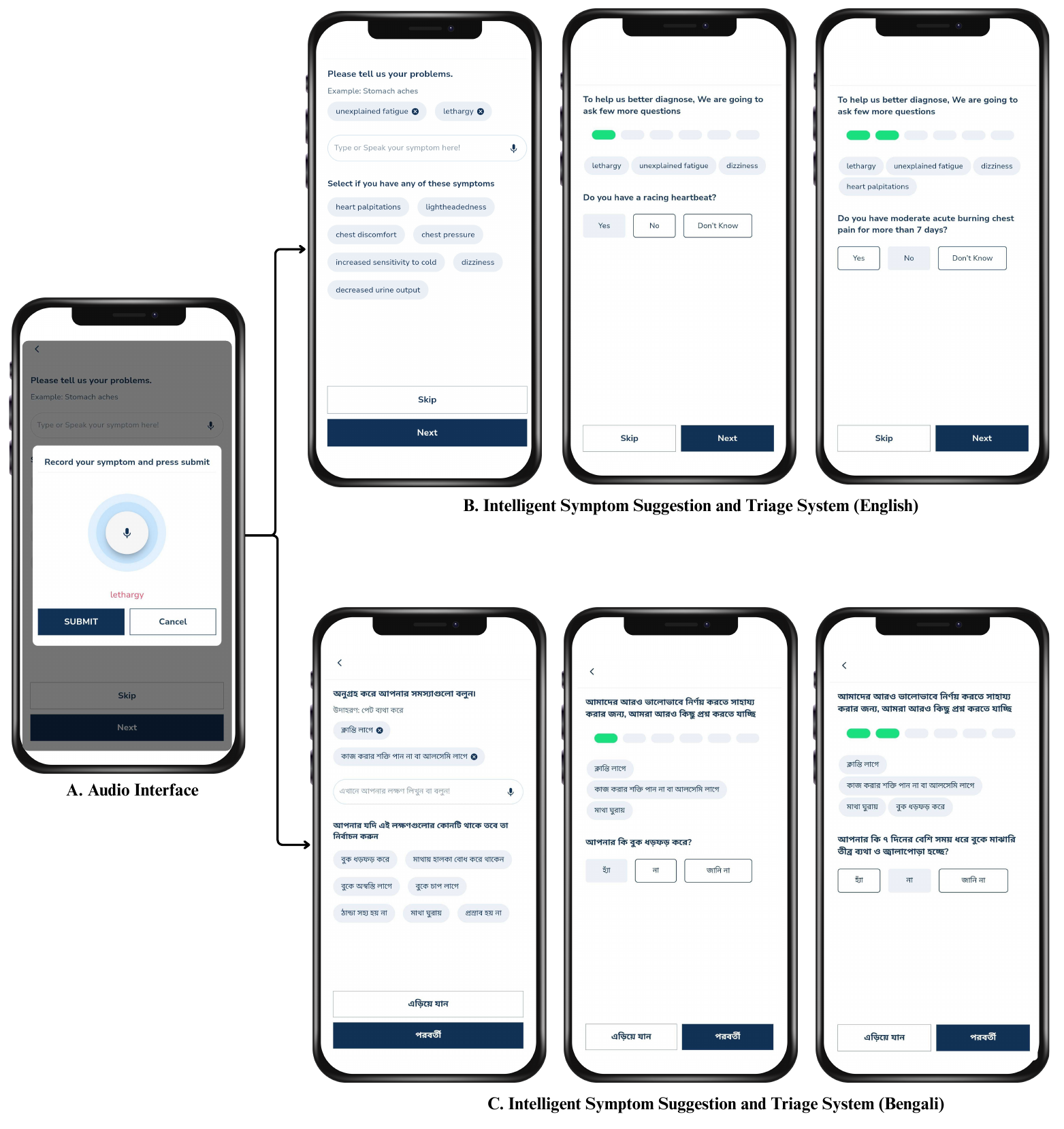}
  \caption{Illustration of intelligent symptom reporting and triage flow: (A) ASR-enabled voice-interactive symptom reporting interface; (B) and (C) show symptom suggestion and dynamic question flow to capture patient's symptoms in English and Bengali.}
  \label{fig:fig11}
\end{figure}
For special symptoms such as pain, fever or cough subsequent sets of questions are prompted following existing clinical guidelines. For example, for pain, seven additional questions are prompted following the \textbf{SOCRATES} (Site, Onset, Character, Radiation, Association, Time Course, Exacerbating / relieving factors and Severity) structured framework to assess pain.
\subsection{Dynamic Symptom Assessment Module}
To optimize the user experience and enhance diagnostic accuracy, the symptom checker integrates a Symptom Suggestion and a Dynamic Assessment module. The Symptom Suggestion module analyzes a user's initial symptom input, employing a sophisticated algorithm to identify potential correlations and propose additional relevant symptoms (\textit{Figure }\ref{fig:fig11}). This functionality is designed to mitigate the common challenge of patients under-reporting symptoms, often due to limited medical knowledge or apprehension. By systematically expanding the symptom list, this module encourages users to provide a more comprehensive representation of their health concerns, thereby significantly improving the effectiveness of the decision-support system.

Following the initial symptom collection, the dynamic symptom assessment module generates a series of tailored questions directly pertinent to the user's reported symptoms. This interactive process facilitates the iterative refinement of the user's health profile. The system's question selection algorithm is inherently adaptive, adjusting subsequent inquiries based on the user's preceding responses. This approach ensures that the generated questions remain highly relevant, engaging, and informative throughout the assessment process. For instance, the system may inquire, "\textbf{Do you have moderate acute burning chest pain for more than 7 days?}" as a tailored follow-up to a positive response regarding a previous symptom, "\textbf{Do you have a racing heartbeat?}" (\textit{Figure }\ref{fig:fig11}). This iterative dialogue continues for a minimum of six iterations, or until the user elects to terminate the process, or a sufficient confidence score is achieved. The confidence score serves as a quantitative metric for the system's certainty in proposing a provisional clinical decision. Upon the conclusion of the questioning phase, the system recommends a relevant medical specialization for further assessment. It is important to note that while this module aids in gathering information crucial for assessing and predicting provisional underlying condition, it does not provide definitive triage decisions or prioritize patients based on urgency; these responsibilities remain with the healthcare provider.

\subsection{Dual-Stream Recommendations: Specialist Guidance for Patients and Provisional Diagnoses for Physicians}

\begin{figure}[!ht]
  \centering
  \includegraphics[width=14cm]{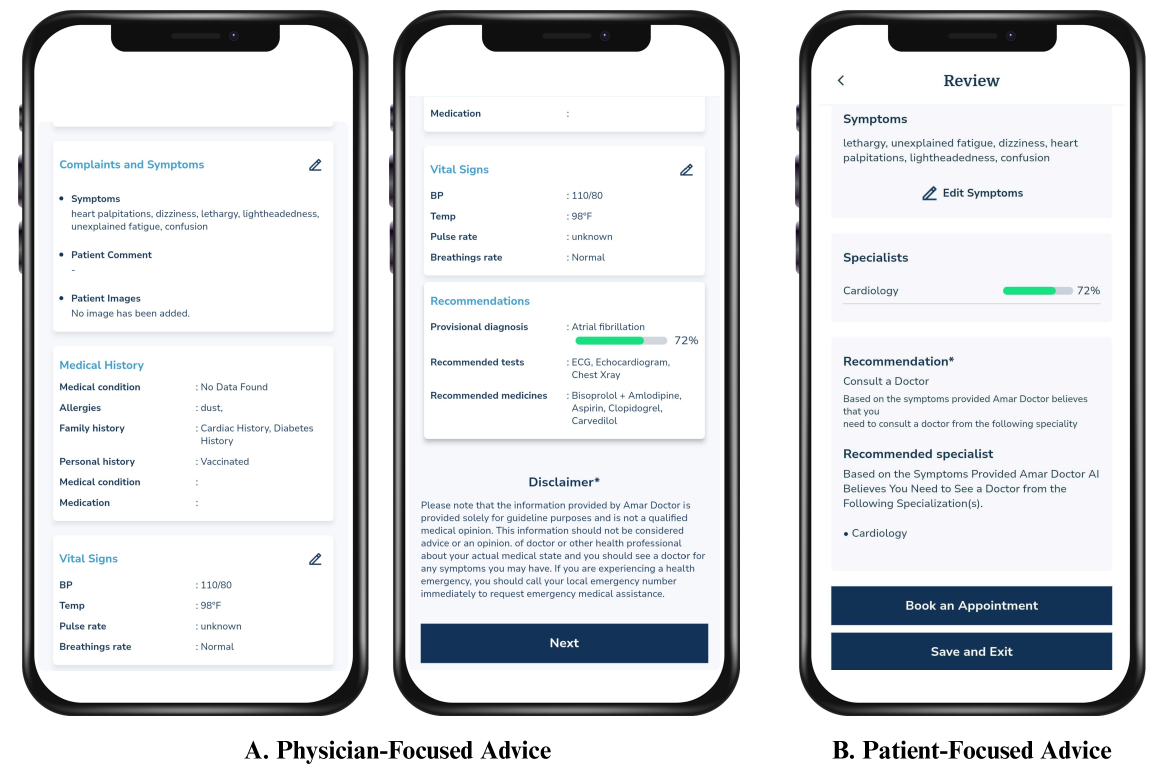}
  \caption{Overview of the final recommendations received by physicians and patients: (\textbf{A}) displays two screens, illustrating structured \textbf{SOAP} note comprising patient's physiological and medical data and provisional recommendation with associated confidence score received by physician (\textbf{B}) displaying specialization recommendation with confidence score visualized in patient's side.}
  \label{fig:fig2}
\end{figure}

Upon completion of the dynamic questionnaire, the system generates two distinct, user-specific recommendations (Figure \ref{fig:fig2}).

For the patient, a recommendation for the appropriate medical specialist type is generated, accompanied by a confidence score associated with the recommendation. This guidance is particularly critical in low- and middle-income countries, such as Bangladesh, where established general practice (GP) or family doctor systems are often absent. In such contexts, patients frequently engage in self-diagnosis and direct self-referral to specialist clinics. Given the prevalent low literacy levels among a significant portion of the patient population, this often leads to misdirected referrals, resulting in wasted time and financial resources, and crucially, delaying accurate diagnosis and timely intervention.

When the patient books an appointment on the platform, physicians are provided with a comprehensive Subjective, Objective, Assessment, and Plan (\textbf{SOAP}) note generated by the system. This note includes: (i) a provisional diagnosis, derived from our proprietary clinical decision support AI model, with an associated confidence score for each provisional disease diagnosis, and (ii) for each provisional diagnosis associated treatment recommendations encompassing medication, dosage, and diagnostic tests (Plan). These elements collectively enable healthcare providers to rapidly assess the patient's potential medical condition. Our platform functions as a machine-in-the-loop assistant, supporting physicians by offering data-driven insights that can highlight areas requiring further investigation or immediate attention.

The core strength of AmarDoctor resides in its secure, intelligent platform, which transforms a simple patient questionnaire into a powerful diagnostic and referral tool. By synergistically combining patient-reported medical concerns with advanced AI recommendations and analytical capabilities, the system fosters a collaborative healthcare experience. This empowers both patients and medical professionals with timely, accurate, and personalized clinical information. Furthermore, the platform's design prioritizes safety and confidentiality, ensuring that all personal health information is processed and transmitted with the highest standards of data protection and privacy.

\subsection{Disease and Symptom Coverage}

\subsubsection{Disease Coverage}

As a primary care management tool, AmarDoctor's investigation encompassed disease categories spanning \textbf{18} distinct medical specializations, including Medicine / General Physician, Cardiology, Neuromedicine, and Gastroenterology (\textit{Supplementary Figure 1}). This focus is particularly pertinent given the pressing need for accessible primary care in many low- and middle-income countries. Our methodology provides an initial clinical assessment, which can inform future treatment strategies and help mitigate disease progression, especially in contexts where timely access to general practitioners is limited.

However, it is crucial to acknowledge the inherent limitations of remote assessment for certain specialized disease classes that necessitate detailed physical examination and in-person physician consultation for accurate differential diagnosis. Consequently, we have deliberately excluded diseases from several specializations, such as \textbf{Neonatology, Oncology, Neoplasms, Surgical emergencies, and Acute conditions}, from the AmarDoctor platform. Similarly, trauma, acute conditions (e.g., burns, severe trauma), or any surgical emergencies have been excluded due to the critical requirement for immediate physical intervention and specialized diagnostic tests. Furthermore, major psychiatric illnesses, such as depression and anxiety, were also excluded. This decision stems from their subjective nature, which can lead to misleading self-assessments by patients and necessitates careful, nuanced evaluation by trained mental health professionals.

\subsubsection{Symptom Coverage}
\begin{figure}[!ht]
    \centering
    \includegraphics[width=13cm, scale=1]{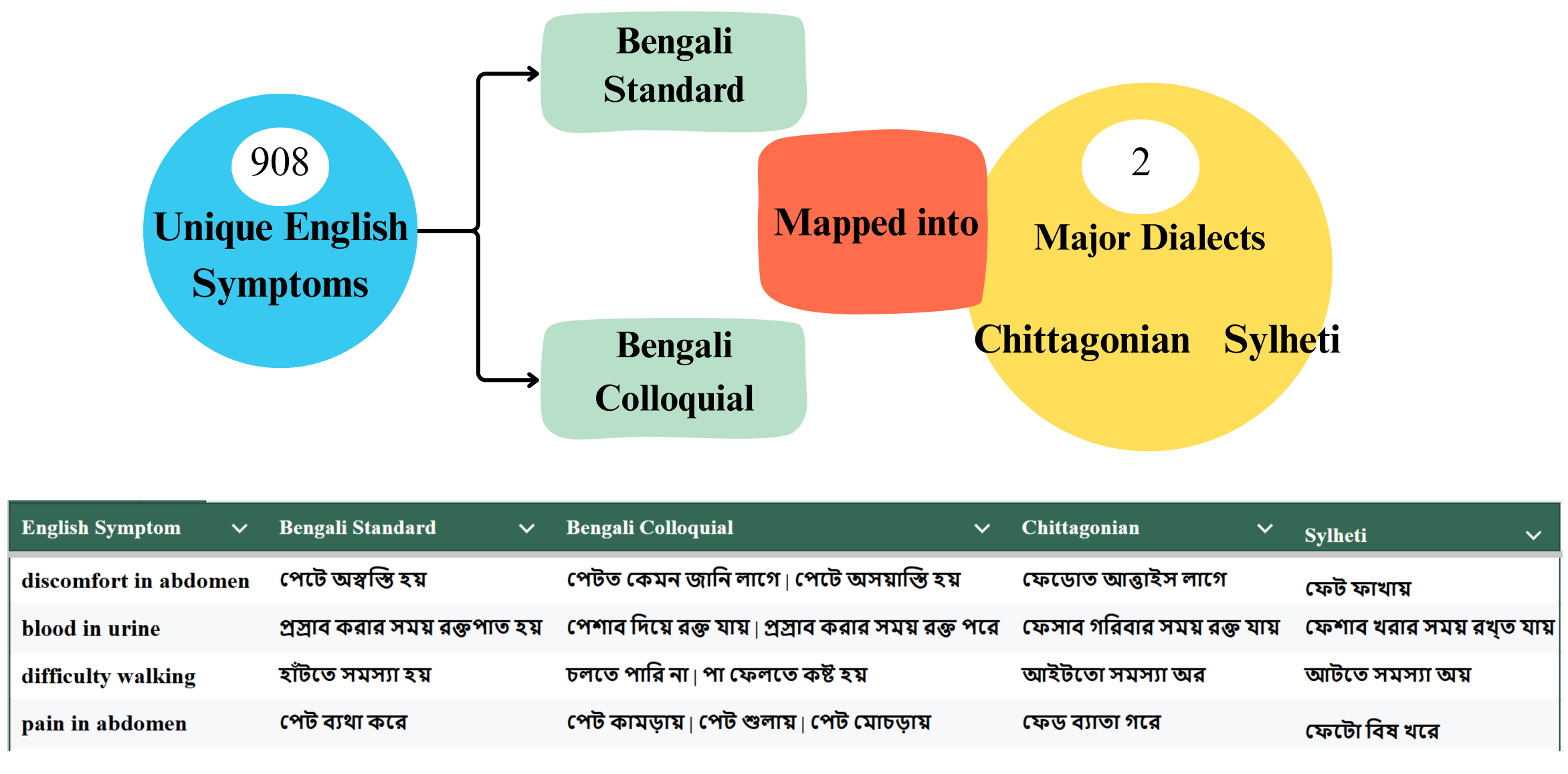}
    \centering
    \caption{Comprehensive symptom mapping workflow: curation of 908 English symptoms through medical team collaboration; followed by translation into standard Bengali, local variations, and Chittagonian and Sylheti dialects to capture linguistic diversity and ensure precise medical communication}
    \label{fig:fig3}
\end{figure}

To facilitate accurate symptom identification, our medical team curated a comprehensive list of \textbf{908} unique English symptoms. These symptoms were meticulously translated into Bengali, incorporating common regional dialects and colloquialisms to ensure linguistic precision. This patient-centered approach enables individuals to articulate their medical concerns using familiar language, as our system is designed to recognize multiple Bengali synonyms for each term. Such an approach is vital for accurate data collection, which, in turn, underpins robust decision support during diagnosis. To further enhance accessibility, translations for two major Bengali dialects, \textbf{Chittagonian} and \textbf{Sylheti}, were also incorporated (\textit{Figure} \ref{fig:fig3}). Collectively, our system now recognizes over \textbf{3,237} Bengali variations, including the original translations. This extensive coverage significantly improves the patient experience by allowing them to express themselves comfortably and access more effective healthcare.

\begin{figure}[!ht]
    \centering
    \includegraphics[width=14cm]{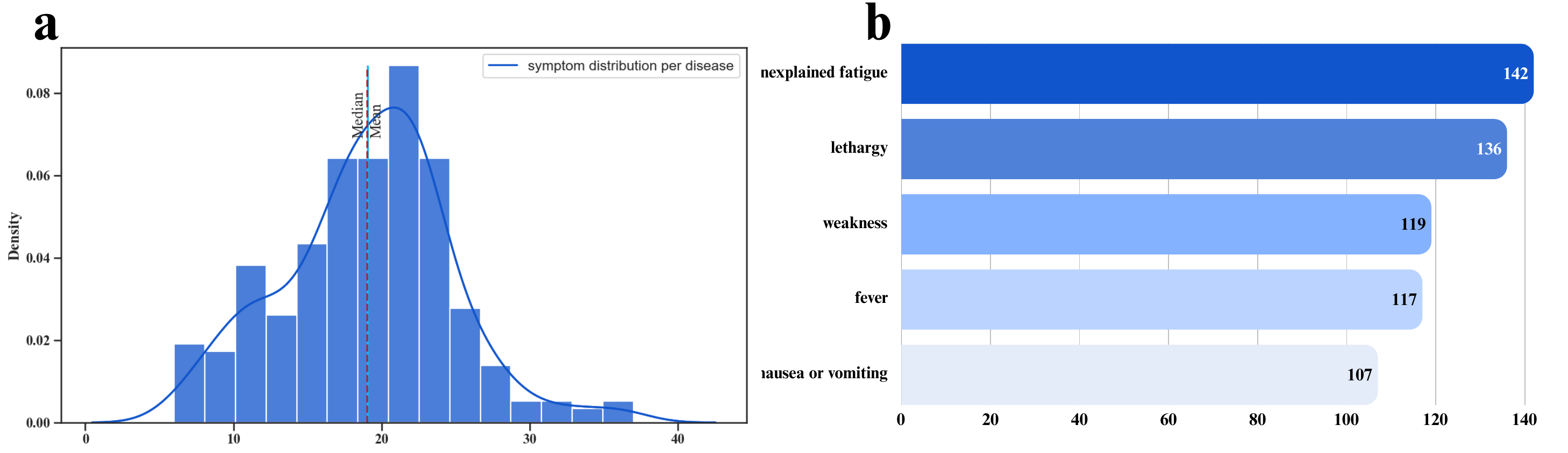}
   \caption{Visualizations of symptom distribution across diseases: \textbf{a}, illustrates the overall distribution, while \textbf{b }highlights the top five most frequently occurring symptoms for each disease.}
   \label{fig:fig12}
\end{figure}

\textit{Figure} \ref{fig:fig12}A provides a comprehensive visualization of how symptoms are distributed among different diseases, showcasing significant variability in symptom occurrence. On average, each disease in the study is characterized by \textbf{19} distinct symptoms, with a notable range spanning from a minimum of \textbf{6} to a maximum of \textbf{37} symptoms. In the dataset, "unexplained fatigue" was the most prevalent symptom, occurring \textbf{142} times \textit{Figure} \ref{fig:fig12}B. 

\subsection{Clinical Vignette Creation for Evaluation of the Recommendation System} 

To establish a robust framework for evaluating the performance of \textbf{AmarDoctor}, we meticulously crafted a set of \textbf{185} patient clinical vignettes (\textit{Supplementary Table 4}), a validated method for assessing clinical quality and diagnostic accuracy \textcolor{blue}{\cite{Peabody2000}}. These patient cases were initially developed by a clinical consultant Sumaiya Tasnia Khan (\textbf{STK}) with \textbf{7} years of experience as a general physician from Central Police Hospital, Bangladesh. Subsequently, two senior clinical experts—Suparna Das (\textbf{SD}), with \textbf{12} years of extensive experience in general practice and emergency care (NHS, England), and Moushumi Paul (\textbf{MP}), with \textbf{11} years of experience in general practice (Obstetrics \& Gynaecology) (RMO, Australia)—independently reviewed and validated these vignettes (\textit{see  Acknowledgements}). In this process, both experts leveraged their extensive clinical experience, insights from pilot study data, and a deep understanding of disease prevalence and risk factors specific to the Bangladeshi patient population, including how symptoms are typically reported in these contexts. This rigorous approach was designed to validate the precision of our recommendation system specifically within the context of the Bangladeshi healthcare system.

\begin{figure}[!ht]
    \centering
    \includegraphics[width=15cm]{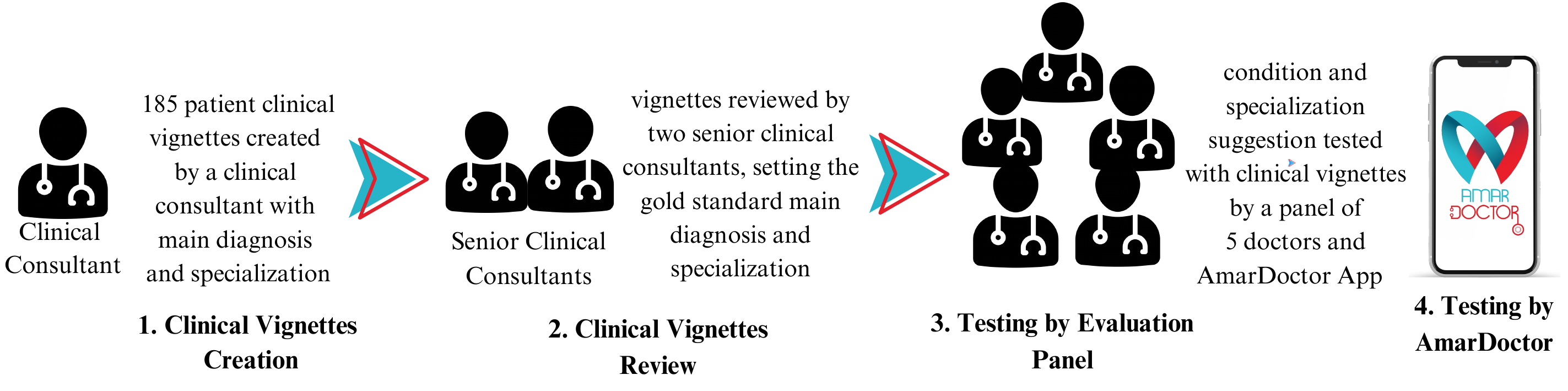}
    \centering
    \caption{Overview of the study methodology including: (1) vignette creation, (2) vignette review, (3) testing by physicians based on diagnosis and specialization recommendation, (4) testing by AmarDoctor}
    \label{fig:fig4}
\end{figure}

The vignettes collectively represent a spectrum of \textbf{103} distinct disease terms (e.g., \textbf{Chronic Kidney Disease}, \textbf{Acute Gastritis} [\textit{Supplementary Table 1}]), which are further grouped under \textbf{78} parent disease terms (e.g., \textbf{Kidney disease}, \textbf{Gastritis}) commonly encountered in general practice. This comprehensive set was refined from an initial pool of \textbf{279} disease terms [\textit{Supplementary Table 2}]. A critical aspect of this curation involved integrating conditions frequently reported among ethnic Bengalis in both the UK and Bangladesh, thereby explicitly addressing the crucial consideration of racial disparity in healthcare data and outcomes. The scope of these patient vignettes was designed to encompass a wide range of disease groups, spanning \textbf{18} distinct medical specializations (\textit{Supplementary Table 3}). Notably, \textbf{"Medicine / General Physician"} emerged as the most prevalent category within this dataset, accounting for \textbf{18.37\%} of the cases. This prevalence underscores the broad spectrum of health concerns typically addressed by general practitioners, who serve as the primary point of contact for a significant portion of the population. The emphasis on "Medicine / General Physician" cases is attributed to factors such as the general nature of many common health conditions, the accessibility of primary care services, and the pivotal role of physicians in providing initial assessments and referrals to specialized care when necessary. While "Medicine / General Physician" was the most common specialization, other areas such as \textbf{"Cardiology"}, \textbf{"Neuromedicine"}, \textbf{"Gastroenterology"}, \textbf{"Rheumatology"} and \textbf{"Respiratory / Chest Disease"} also represented a considerable portion of the disease groups. This diversity reflects the complexity of human health and the need for specialized expertise in various medical domains.

Each vignette was meticulously designed to encompass comprehensive patient data, including age, medical history (e.g., pregnancy, heart disease, hypertension, diabetes), family history (e.g., history of diabetes, heart disease, hypertension in the family), chief complaints, and additional symptoms. This detailed design covered both common and rare conditions pertinent to primary care practice, incorporating diverse clinical presentations and conditions affecting all body systems, thus creating comprehensive and realistic scenarios for analysis. These vignettes were carefully constructed from examples of multiple real patient cases to ensure close proximity to authentic patient encounters while maintaining anonymity and preventing traceability. To ensure dataset diversity, a balanced gender distribution (\textbf{51\% male}, \textbf{49\% female}) and an age range of \textbf{18} to \textbf{78} years were maintained. Furthermore, each vignette was assigned a gold-standard main diagnosis and corresponding physician's specialization, as definitively determined by our clinical expert \textbf{STK} and senior clinical practitioners \textbf{SD} and \textbf{MP}. This rigorous, multi-stage peer-review process, informed by real-world clinical context, ensured the high accuracy and clinical relevance of our vignettes, establishing a solid and credible foundation for the subsequent comparative analysis of the symptom checker applications.

\begin{table}[!ht]
\begin{tabular}{|l|l|}
\hline
\textbf{Patient ID}                                                                                            & \textbf{Patient 12}                                                                                                                                                        \\ \hline
\textbf{Sex}                                                                                                   & Female                                                                                                                                                                     \\ \hline
\textbf{Age}                                                                                                   & 30                                                                                                                                                                         \\ \hline
\textbf{Family History}                                                                                        & ......                                                                                                                                                                     \\ \hline
\textbf{Medical History}                                                                                       & ......                                                                                                                                                                     \\ \hline
\textbf{Any Current Medication}                                                                                & ......                                                                                                                                                                     \\ \hline
\textbf{Allergies}                                                                                             & Dust, Pets                                                                                                                                                                 \\ \hline
\textbf{Any Other Remarks}                                                                                     & History of exposure to allergens                                                                                                                                           \\ \hline
\textbf{Primary Complaints}                                                                                    & \begin{tabular}[c]{@{}l@{}}wheezing,\\ dry cough,\\ mild fever\end{tabular}                                                                                                \\ \hline
\textbf{Additional Symptoms}                                                                                   & \begin{tabular}[c]{@{}l@{}}shortness of breath at rest, \\ persistent cough, \\ blue or gray skin color due to low oxygen levels, \\ acute episodes, lethargy\end{tabular} \\ \hline
\textbf{\begin{tabular}[c]{@{}l@{}}Diagnosis 1, \\ Diagnosis 2, \\ Diagnosis 3\end{tabular}}                   & \begin{tabular}[c]{@{}l@{}}Flu,\\ Asthma, \\ Upper respiratory tract infection\end{tabular}                                                                                \\ \hline
\textbf{\begin{tabular}[c]{@{}l@{}}Medication 1,\\ Medication 2, \\ Medication 3\end{tabular}}                 & \begin{tabular}[c]{@{}l@{}}Salbutamol, \\ Montelukast Sodium 10mg,\\ paracetamol 500 mg\end{tabular}                                                                       \\ \hline
\textbf{\begin{tabular}[c]{@{}l@{}}Diagnostic Test 1, \\ Diagnostic Test 2, \\ Diagnostic Test 3\end{tabular}} & \begin{tabular}[c]{@{}l@{}}Chest Xray (P/A view), \\ CBC, \\ MT test\end{tabular}                                                                                          \\ \hline
\textbf{Advice}                                                                                                & Home care        
       \\ \hline
\textbf{Specialization}                                                                                                & General Physician / Medicine

\\ \hline
\textbf{\begin{tabular}[c]{@{}l@{}}Your Reason Behind Diagnosis, \\ Additional Comments\end{tabular}}          & ......                                                                                                                                                                     \\ \hline
\end{tabular}
\centering
\vspace{0.12cm}
\caption{Comprehensive patient profile and clinical assessment for vignette patient 12, encompassing demographic data, medical history, current health status, and physician-provided clinical evaluation}
\label{tab:tab1}
\end{table}
\subsection{Independent Evaluation of The Recommendation System using External Physicians}

To evaluate the performance of our platform, five external physicians were tasked with carefully reviewing a series of clinical vignettes (\textit{Table} \ref{tab:tab1}). These physicians, registered and licensed by the Bangladesh Medical and Dental Council (\textbf{BMDC}), possessed an average of three years of clinical experience. Each physician was required to examine all the \textbf{185} vignettes and make provisions on the top three provisional diagnoses, suggest specialization, offer advice on the level of urgency, and explain the rationale behind their judgments, along with any additional comments.

This expert evaluation provided valuable insights into the symptom checker's ability to accurately identify potential health concerns and guide patients toward appropriate medical interventions. By comparing the physicians' diagnoses and recommendations to the system's outputs, we were able to assess the system's effectiveness in replicating the decision-making processes of experienced healthcare professionals.
\subsection{Comparative Evaluation of AI and Human Diagnostic Performance}
To comprehensively evaluate the accuracy of AmarDoctor's AI generated provisional disease diagnosis and specialization recommendation, we employed a rigorous comparison strategy. This involved assessing the provisional diagnoses and specialization suggestions provided by both our system and a panel of five external physicians against a predefined gold-standard main diagnosis on the patient vignettes. For condition suggestion accuracy, two key metrics were utilized as described in Figure 7.
\begin{figure}[!ht]
\centering
\includegraphics[width=16cm]{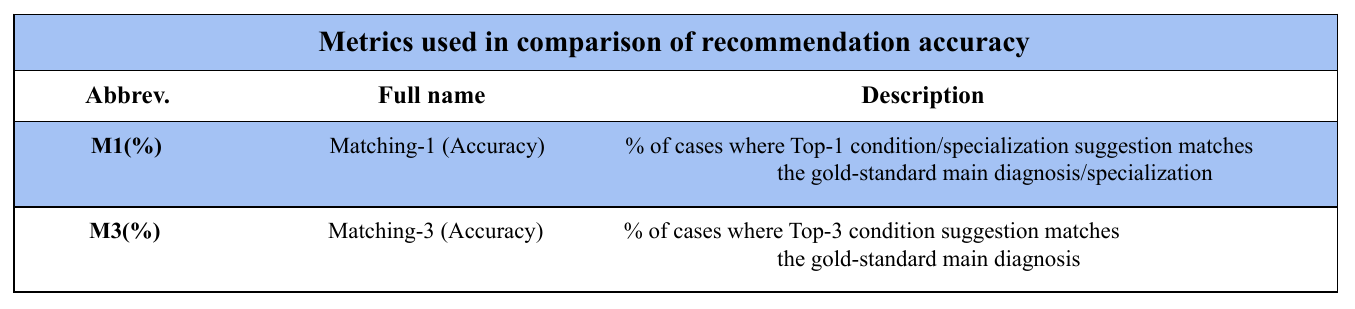}
\centering
\caption{Provisional Disease Diagnosis comparison strategy: M1 - percentage of cases with top recommendation matching main diagnosis, M3 - percentage of cases where top three recommendations contain the main diagnosis}
\label{fig:fig6}
\end{figure}
For provisional disease diagnosis both M1 and M3 metrics was used and for specialization recommendation comparison with physicians, only the \textbf{M1} metric was applied. This approach allowed for a direct assessment of whether the top suggested specialization aligned with the gold standard. A detailed description of these metrics, including their application for both condition and specialization recommendation accuracy, is presented in \textit{Figure} \ref{fig:fig6}. This systematic comparison enabled us to quantify the agreement and discrepancies between the AI-driven recommendations and human clinical judgment, providing a robust measure of our system's performance and its potential utility in a real-world clinical setting.

\section{RESULTS}
Here we present the findings from the comprehensive evaluation of the AmarDoctor platform's diagnostic and specialization recommendation capabilities. The results are derived from an independent review of \textbf{185} patient clinical vignettes by five external physicians, whose assessments are quantitatively compared against AmarDoctor's outputs using the defined \textbf{M1} and \textbf{M3} metrics (as detailed in \textit{Figure} \ref{fig:fig6}). These findings highlight the system's performance in generating provisional diagnoses and suggesting appropriate specializations.

\subsection{Accuracy of Provisional Diagnosis Recommendation}

\begin{figure}[!ht]
    \centering
    \includegraphics[width=15cm]{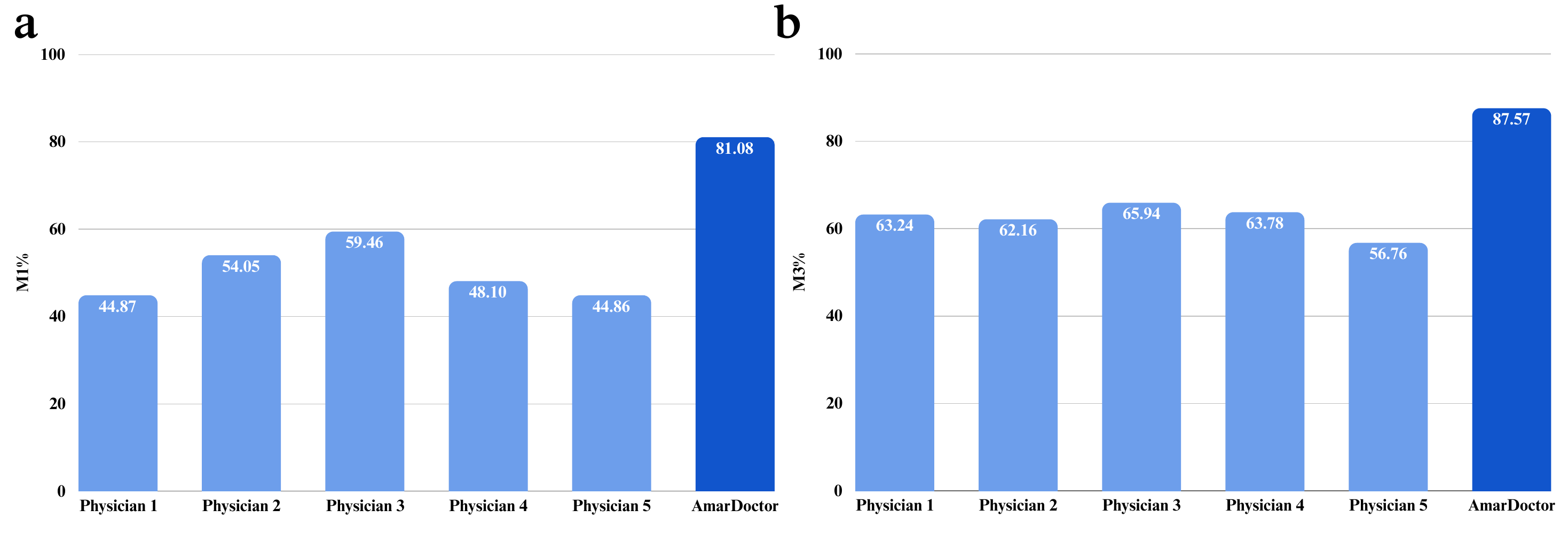}
    \caption{A comparative analysis of condition suggestion accuracy, evaluated against the assessments of panel physicians.  (\textbf{a}), presents the results obtained using the M1 metric, while (\textbf{b}) displays the results achieved using the M3 metric.}
    \label{fig:m1_m3}
   
\end{figure}

The comparative analysis of the results revealed a notable difference in provisional diagnosis recommendation performance between the AmarDoctor AI recommendation system and the panel physicians. In the specific context of virtual medical consultations, where diagnostic reliance is primarily on patient-reported symptoms, AmarDoctor's AI recommendation system demonstrated a higher level of accuracy compared to the average performance observed among the panel physicians. Quantitatively, AmarDoctor achieved an \textbf{81.08\%} accuracy for the M1 metric and an \textbf{87.57\%} for the M3 metric. Conversely, the average accuracies recorded for the participating physicians stood at \textbf{50.27\%} for M1 and \textbf{62.27\%} for M3 (\textit{Figure} \ref{fig:m1_m3}).

\begin{table}[!ht]
\resizebox{\columnwidth}{!}{
\begin{tabular}{|l|l|l|l|l|l|l|l|l|l|l|}
\hline
\textbf{Patients} & \textbf{\begin{tabular}[c]{@{}l@{}}Gold Standard \\ Main Diagnosis\end{tabular}} & \textbf{Physician\_1}                                                                  & \textbf{Physician\_2}                                                                  & \textbf{Physician\_3}                                                               & \textbf{Physician\_4}                                                                & \textbf{Physician\_5}                                                    & \textbf{Our Suggestion}                                                                                         & \textbf{M1} & \textbf{M3} & \textbf{\begin{tabular}[c]{@{}l@{}}Physician \\ Match\end{tabular}} \\ \hline
Patient 37        & Respiratory infection                                                            & \begin{tabular}[c]{@{}l@{}}Bronchitis\\ Asthma\\ Flu\end{tabular}                   & \begin{tabular}[c]{@{}l@{}}Asthma\\ Bronchitis\\ Respiratory infection\end{tabular} & \begin{tabular}[c]{@{}l@{}}Anemia\\ Asthma\\ -\end{tabular}                      & \begin{tabular}[c]{@{}l@{}}Angina\\ Bronchitis\\ Asthma\end{tabular}              & \begin{tabular}[c]{@{}l@{}}COPD\\ Bronchitis\\ -\end{tabular}         & \begin{tabular}[c]{@{}l@{}}Asthma 4/5\\ Asthma 4/5\\ Heart disease 0/5\end{tabular}                             & failure     & failure     & success                                                          \\ \hline
Patient 227       & Gout                                                                             & \begin{tabular}[c]{@{}l@{}}Gout\\ Rheumatoid arthritis\\ Arthritis\end{tabular}     & \begin{tabular}[c]{@{}l@{}}Gout\\ -\\ -\end{tabular}                                & \begin{tabular}[c]{@{}l@{}}Arthritis\\ Osteoarthritis\\ -\end{tabular}           & \begin{tabular}[c]{@{}l@{}}Gout\\ Arthritis\\ -\end{tabular}                      & \begin{tabular}[c]{@{}l@{}}Gout\\ Arthritis\\ -\end{tabular}          & \begin{tabular}[c]{@{}l@{}}Arthritis 4/5\\ Arthritis 4/5\\ Arthritis 4/5\end{tabular}                           & failure     & failure     & success                                                          \\ 
\hline
Patient 253       & Renal disorder                                                                   & \begin{tabular}[c]{@{}l@{}}Diabetes mellitus\\ Hypertension\\ -\end{tabular}        & \begin{tabular}[c]{@{}l@{}}Diabetes mellitus\\ -\\ -\end{tabular}                   & \begin{tabular}[c]{@{}l@{}}Diabetes mellitus\\ -\\ -\end{tabular}                & \begin{tabular}[c]{@{}l@{}}Diabetes mellitus\\ Diabetes mellitus\\ -\end{tabular} & \begin{tabular}[c]{@{}l@{}}Diabetes mellitus\\ -\\ -\end{tabular}     & \begin{tabular}[c]{@{}l@{}}Diabetes mellitus 5/5\\ Diabetes mellitus 5/5\\ Diabetic oculopathy 0/5\end{tabular} & failure     & failure     & success                                                          \\ \hline
Patient 420       & Anemia                                                                           & \begin{tabular}[c]{@{}l@{}}Malaria\\ Hepatitis\\ Meningitis\end{tabular}            & \begin{tabular}[c]{@{}l@{}}Liver disease\\ -\\ -\end{tabular}                       & \begin{tabular}[c]{@{}l@{}}Liver disease\\ -\\ -\end{tabular}                    & \begin{tabular}[c]{@{}l@{}}Viral fever\\ Hepatitis\\ -\end{tabular}               & \begin{tabular}[c]{@{}l@{}}Hepatitis\\ Cholangitis\\ -\end{tabular}   & \begin{tabular}[c]{@{}l@{}}Liver disease 2/5\\ Hepatitis 3/5\\ Hepatitis 3/5\end{tabular}                       & failure     & failure     & success                                                          \\ \hline
Patient 444       & Heart disease                                                                           & \begin{tabular}[c]{@{}l@{}}Heart disease\\ Heart failure\\ Angina\end{tabular}            & \begin{tabular}[c]{@{}l@{}}Cardiac arrest\\ -\\ -\end{tabular}                       & \begin{tabular}[c]{@{}l@{}}Congestive heart failure\\ Angina\\ -\end{tabular}                    & \begin{tabular}[c]{@{}l@{}}Cardiac arrest\\ Heart disease\\ -\end{tabular}               & \begin{tabular}[c]{@{}l@{}}Angina\\ Myocardial infarction\\ Atrial fibrillation\end{tabular}   & \begin{tabular}[c]{@{}l@{}}Myocardial infarction 1/5\\ Angina 3/5\\ Heart failure 1/5\end{tabular}                       & failure     & failure     & success                                                          \\ \hline
Patient 448       & Endometriosis                                                                           & \begin{tabular}[c]{@{}l@{}}Pelvic inflammatory disease\\ Endometriosis\\ Prolapse\end{tabular}            & \begin{tabular}[c]{@{}l@{}}Pelvic inflammatory disease\\ -\\ -\end{tabular}                       & \begin{tabular}[c]{@{}l@{}}Adenomyosis\\ -\\ -\end{tabular}                    & \begin{tabular}[c]{@{}l@{}}Endometriosis\\ Adenomyosis\\ -\end{tabular}               & \begin{tabular}[c]{@{}l@{}}Pelvic inflammatory disease\\ Endometriosis\\ Uterine fibroid\end{tabular}   & \begin{tabular}[c]{@{}l@{}}Pelvic inflammatory disease 3/5\\ Pelvic inflammatory disease 3/5\\ Inflammatory bowel disease 0/5\end{tabular}                       & failure     & failure     & success                                                          \\ \hline
\end{tabular}
} 
\vspace{2pt}
\caption{ Selected vignette cases where AmarDoctor's diagnostic recommendations diverged from gold standard yet achieved physician consensus (6 of 10 cases)}
\label{tab:vignette1}
\end{table}
\begin{table}[!ht]
\resizebox{\columnwidth}{!}{
\begin{tabular}{|l|l|l|l|l|l|l|l|l|l|l|}
\hline
\textbf{Patients} & \textbf{\begin{tabular}[c]{@{}l@{}}Gold Standard \\ Main Diagnosis\end{tabular}} & \textbf{Physician\_1}                                                                  & \textbf{Physician\_2}                                                                  & \textbf{Physician\_3}                                                               & \textbf{Physician\_4}                                                                & \textbf{Physician\_5}                                                    & \textbf{Our Suggestion}                                                                                         & \textbf{M1} & \textbf{M3} & \textbf{\begin{tabular}[c]{@{}l@{}}Physician \\ Match\end{tabular}} \\ \hline
Patient 12        & Asthma                                                                           & \begin{tabular}[c]{@{}l@{}}Flu\\ Asthma\\ Respiratory infection\end{tabular}        & \begin{tabular}[c]{@{}l@{}}Bronchitis\\ Asthma\\ Respiratory infection\end{tabular} & \begin{tabular}[c]{@{}l@{}}Asthma\\ -\\ -\end{tabular}                           & \begin{tabular}[c]{@{}l@{}}Asthma\\ Bronchitis\\ -\end{tabular}                   & \begin{tabular}[c]{@{}l@{}}Asthma\\ Bronchitis\\ -\end{tabular}       & \begin{tabular}[c]{@{}l@{}}Heart failure 0/5\\ Heart failure 0/5\\ Heart failure 0/5\end{tabular}               & failure     & failure     & failure                                                          \\ \hline
Patient 131       & COPD                                                                             & \begin{tabular}[c]{@{}l@{}}Asthma\\ Bronchitis\\ Respiratory infection\end{tabular} & \begin{tabular}[c]{@{}l@{}}Respiratory infection\\ -\\ -\end{tabular}               & \begin{tabular}[c]{@{}l@{}}Respiratory infection\\ -\\ -\end{tabular}            & \begin{tabular}[c]{@{}l@{}}Asthma\\ Respiratory infection\\ -\end{tabular}        & \begin{tabular}[c]{@{}l@{}}COPD\\ Bronchitis\\ Asthma\end{tabular}    & \begin{tabular}[c]{@{}l@{}}Heart failure 0/5\\ Heart failure 0/5\\ Heart failure 0/5\end{tabular}               & failure     & failure     & failure                                                          \\ \hline

Patient 132       & Asthma                                                                           & \begin{tabular}[c]{@{}l@{}}Asthma\\ Bronchitis\\ Respiratory infection\end{tabular}            & \begin{tabular}[c]{@{}l@{}}Pleural effusion\\ -\\ -\end{tabular}                       & \begin{tabular}[c]{@{}l@{}}Respiratory infection\\ Tuberculosis\\ -\end{tabular}                    & \begin{tabular}[c]{@{}l@{}}Bronchitis\\ Asthma\\ -\end{tabular}               & \begin{tabular}[c]{@{}l@{}}Asthma\\ Bronchitis\\ Chronic obstructive pulmonary disease\end{tabular}   & \begin{tabular}[c]{@{}l@{}}Heart failure 0/5\\ Heart failure 0/5\\ Heart failure 0/5\end{tabular}                       & failure     & failure     & failure                                                          \\ \hline
\end{tabular}
}
\vspace{2pt}
\caption{Cases where AmarDoctor's condition suggestions diverged from the gold standard main diagnosis and lacked support from physician review.}
\label{tab:vignette2}
\end{table}
\begin{table}[!ht]
\resizebox{\columnwidth}{!}{
\begin{tabular}{|l|l|l|l|l|l|l|l|l|l|l|}
\hline
\textbf{Patients} & \textbf{\begin{tabular}[c]{@{}l@{}}Gold Standard \\ Main Specialization\end{tabular}} & \textbf{Physician\_1}           & \textbf{Physician\_2}           & \textbf{Physician\_3}            & \textbf{Physician\_4}          & \textbf{Physician\_5}                                                     & \textbf{Our Suggestion}                                                                                                            & \textbf{M1} & \textbf{M3} & \textbf{\begin{tabular}[c]{@{}l@{}}Physician\\ Match\end{tabular}} \\ \hline
Patient 12        & Respiratory / Chest Disease                                                           & Respiratory / Chest Disease  & Respiratory / Chest Disease  & Respiratory / Chest Disease   & Respiratory / Chest Disease & \begin{tabular}[c]{@{}l@{}}Respiratory / \\ Chest Disease\end{tabular} & \begin{tabular}[c]{@{}l@{}}Cardiology 0/5\\ Cardiology 0/5\\ Cardiology 0/5\end{tabular}                                           & failure     & failure     & failure                                                         \\ \hline
Patient 131       & Respiratory / Chest Disease                                                           & Respiratory / Chest Disease  & Medicine / General Physician & Respiratory / Chest Disease   & Respiratory / Chest Disease & Respiratory / Chest Disease                                            & \begin{tabular}[c]{@{}l@{}}Cardiology 0/5\\ Cardiology 0/5\\ Cardiology 0/5\end{tabular}                                           & failure     & failure     & failure                                                         \\ \hline
Patient 132       & Respiratory / Chest Disease                                                           & Respiratory / Chest Disease  & Medicine / General Physician & Respiratory / Chest Disease   & Respiratory / Chest Disease & Respiratory / Chest Disease                                            & \begin{tabular}[c]{@{}l@{}}Cardiology 0/5\\ Cardiology 0/5\\ Cardiology 0/5\end{tabular}                                           & failure     & failure     & failure                                                         \\ \hline
Patient 253       & Nephrology                                                                            & Diabetes / Endocrinology     & Diabetes / Endocrinology     & Diabetes / Endocrinology      & Diabetes / Endocrinology    & \begin{tabular}[c]{@{}l@{}}Diabetes / \\ Endocrinology\end{tabular}    & \begin{tabular}[c]{@{}l@{}}Diabetes / Endocrinology 5/5\\ Diabetes / Endocrinology 5/5\\ Diabetes / Endocrinology 5/5\end{tabular} & failure     & failure     & success                                                         \\ \hline
Patient 483       & Respiratory / Chest Disease                                                           & ENT                          & ENT                          & Medicine / General Physician  & Respiratory / Chest Disease & Respiratory / Chest Disease                                            & \begin{tabular}[c]{@{}l@{}}Medicine / General Physician 1/5\\ ENT 2/5\\ Hepatology 0/5\end{tabular}                                & failure     & failure     & success                                                         \\ \hline
Patient 420       & \begin{tabular}[c]{@{}l@{}}Haematology | \\ Medicine / General Physician\end{tabular} & Medicine / General Physician & Hepatology                   & Hepatology                    & Hepatology                  & Gastroenterology                                                       & \begin{tabular}[c]{@{}l@{}}Hepatology 3/5\\ Hepatology 3/5\\ Hepatology 3/5\end{tabular}                                           & failure     & failure     & success                                                         \\ \hline
Patient 421       & Hepatology                                                                            & Hepatology                   & Medicine / General Physician & Hepatology                    & Hepatology                  & Gastroenterology                                                       & \begin{tabular}[c]{@{}l@{}}Nephrology 0/5\\ Nephrology 0/5\\ Nephrology 0/5\end{tabular}                                           & failure     & failure     & failure                                                         \\ \hline
\end{tabular}
}
\centering
\vspace{2pt}
\caption{Instances where AmarDoctor's specialization recommendation diverged from the gold standard main specialization}
\label{tab:sp}
\end{table}

\subsection{Analysis of Shared Diagnostic Discrepancies and Complex Cases}
A deeper examination of the diagnostic outcomes reveals specific instances where both the AmarDoctor system and the human clinicians faced similar challenges, particularly in complex cases leading to shared discrepancies with the gold standard. Our evaluation of the \textbf{M3} metric identified a total of \textbf{23} failure cases out of \textbf{185} cases. Among these failures, \textbf{20} instances exhibited concordance between our diagnoses and those of physicians, despite diverging from the gold standard main diagnosis. This discrepancy underscores the inherent challenges in achieving universally accurate diagnoses based solely on the symptom data. Notably, in \textbf{10} of these cases (\textit{Table} \ref{tab:vignette1}), at least one of our top three diagnostic predictions aligned with the assessments from three or more physicians, indicating a degree of shared clinical reasoning.

A detailed examination of specific cases further illuminates the complexities of differential diagnosis. For instance, in \textbf{Patient 37}'s case, while the gold standard diagnosis was "\textbf{Respiratory infection}", both AmarDoctor and four physicians suggested "\textbf{Asthma}". This misclassification is potentially attributable to the considerable symptom overlap between these conditions. Similar diagnostic challenges were observed in other cases: the system's classification of "\textbf{Gout}" as "\textbf{Arthritis}" (\textbf{Patient 227}), "\textbf{Anemia}" as "\textbf{Liver disease}" (noting that hepatic dysfunction commonly presents with hematological abnormalities), "\textbf{Heart disease}" as "\textbf{Myocardial infarction}", and "\textbf{Endometriosis}" as "\textbf{Pelvic inflammatory disease}". These instances collectively underscore the inherent complexity of differential diagnosis and suggest that comprehensive diagnostic accuracy may necessitate the integration of additional clinical data, including detailed patient history, physical examination findings, and targeted diagnostic testing.

A particularly noteworthy case was \textbf{Patient 253}, where the gold standard diagnosis of "\textbf{Renal disorder due to type 1 diabetes mellitus}" (Renal disorder) was classified as "\textbf{Type 1 diabetes mellitus}" (Diabetes mellitus) by both the system and all five physicians. This case vividly illustrates the challenges in differentiating between primary conditions and their secondary complications, as diabetic symptoms often precede and overlap with subsequent kidney complications, requiring careful attention to subtle clinical indicators for precise diagnosis.

\subsection{Analysis of Unique Diagnostic Discrepancies and Metric Utility}

Beyond the shared diagnostic challenges, this subsection scrutinizes cases where the AmarDoctor system's provisional diagnoses diverged uniquely from both the gold standard and the consensus of the physician panel, further illustrating the utility of the M3 metric in capturing broader diagnostic relevance. Among the \textbf{23} discrepant cases, \textbf{3} instances (Patients \textbf{12}, \textbf{131}, and \textbf{132}) (\textit{Table} \ref{tab:vignette2}) involved two \textbf{Asthma} and one \textbf{Chronic obstructive pulmonary disease (COPD)} diagnoses. These were misclassified as Heart failure with no physician support, likely due to overlapping respiratory symptoms such as dyspnea, wheezing, coughing, and respiratory distress. This further underlines the diagnostic challenges in differentiating between cardiac and pulmonary pathologies that present with similar respiratory manifestations.

Overall, the \textbf{M3} metric's ability to identify instances where our diagnoses align with expert opinions, despite deviations from the gold standard, demonstrates the system's potential to provide valuable insights for clinical decision-making. While the \textbf{M3} metric may not be a perfect measure of diagnostic accuracy, it serves as a valuable tool for evaluating the performance of symptom-based diagnostic systems.

\subsection{Accuracy of Specialization Suggestion}
\begin{figure}[!ht]
    \centering
    \includegraphics[width=8cm]{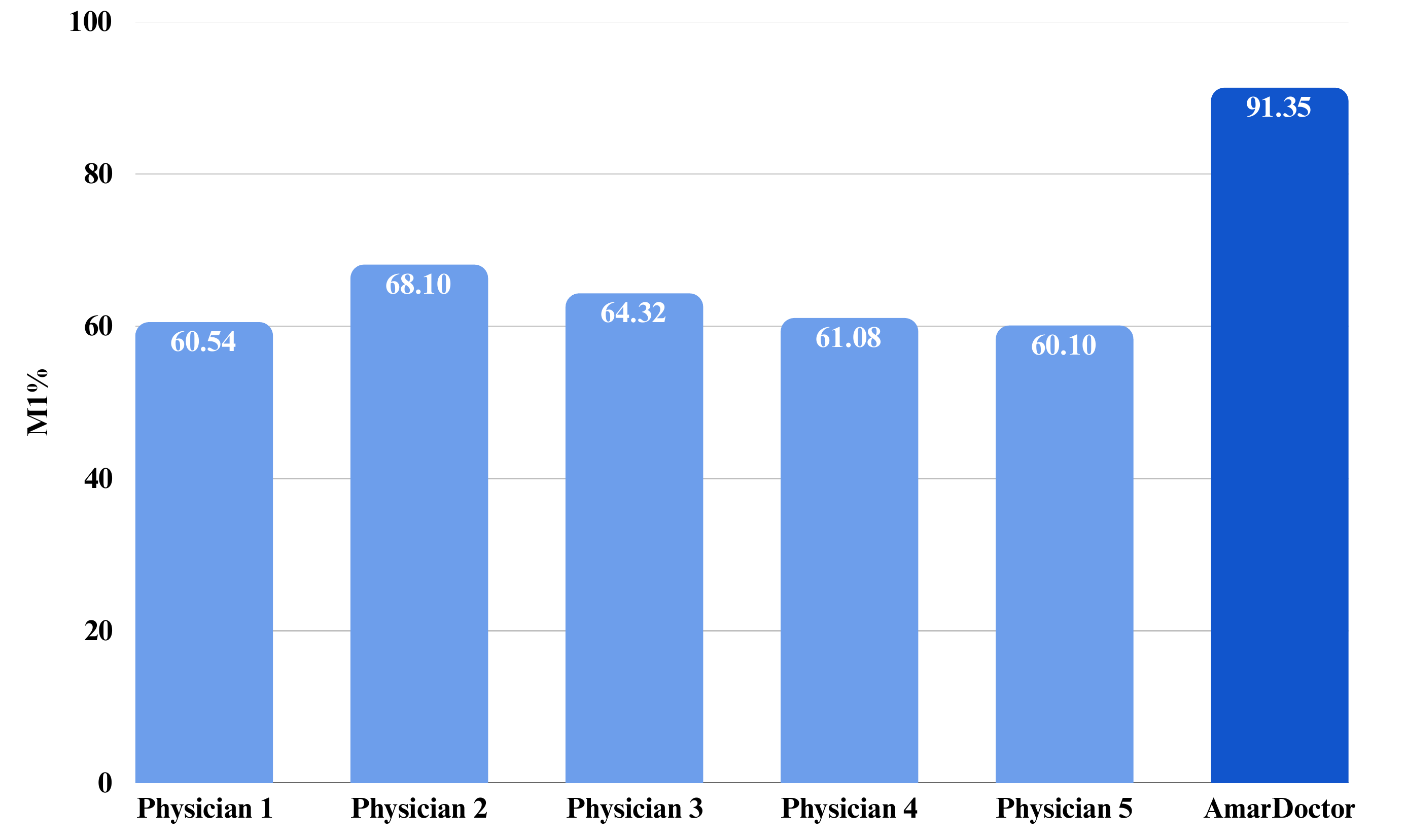}
    \caption{Comparison of M1 Specialization accuracy with the assessments of panel physicians}
    \label{fig:s1}
\end{figure}

Similarly, in terms of specialization recommendation, AmarDoctor outperformed the average physician accuracy of \textbf{62.6\%} by achieving \textbf{91.35\%} accuracy in \textbf{M1} metric (\textit{Figure} \ref{fig:s1}), with \textbf{169} out of \textbf{185} cases correctly identified. The \textbf{M1} metric encountered challenges in \textbf{16} cases, leading to inappropriate specialization suggestions.
While the \textbf{M3} metric revealed \textbf{7} instances of misaligned specialization suggestions that deviated from the primary diagnosis where \textbf{3} of these instances were supported by the consensus of two or more physicians. Most notably, for patient \textbf{253} (as discussed in the previous section), all five reviewing physicians supported the system's recommendation. Similarly, the system's suggestions for patients \textbf{483} and \textbf{420} were endorsed by two and three physicians respectively. This pattern suggests that the system's recommendations often align with expert opinions, even if they deviate from the gold standard.
On the contrary, there were \textbf{4} instances (\textbf{Patient} \textbf{12}, \textbf{131}, \textbf{132}, and \textbf{421}) (\textit{Table} \ref{tab:sp}) where our recommendations were inaccurate and lacked any consensus among the evaluators.

These findings highlight the strengths and limitations of the system's specialization recommendations. While it demonstrates promising performance in many cases, there are instances where further refinement is needed to improve accuracy and align more closely with expert consensus
\section{DISCUSSION}

This article presents AmarDoctor, a pioneering AI-powered digital healthcare platform that introduces the first multilingual symptom checker and clinical decision support system specifically tailored for Bengali, a low-resource language. Our findings demonstrate that AmarDoctor offers a promising and relatively safer approach by functioning as a machine-in-the-loop tool, rather than an autonomous agent, for disease diagnosis and specialization recommendation. This design is particularly pertinent given the severely constrained availability of healthcare providers in developing South Asian nations. By offering structured \textbf{SOAP} (Subjective, Objective, Assessment, and Plan) formatted provisional recommendations, the platform can significantly alleviate the diagnostic burden on medical staff, thereby enhancing the overall consultation experience for both patients and physicians. AmarDoctor’s commitment to accessibility and inclusivity is evident in its multimodal approach, incorporating both text and audio formats, which facilitates a complete voice-interactive triaging process especially beneficial for individuals with low literacy levels.

The empirical evaluation presented in the Result section underscores AmarDoctor's potential to augment diagnostic capabilities in resource-constrained environments. In the specific context of virtual medical consultations, where diagnostic reliance is primarily on patient-reported symptoms, AmarDoctor's AI recommendation system demonstrated a higher level of accuracy compared to the average performance observed among the panel physicians. Quantitatively, AmarDoctor achieved an \textbf{81.08\%} accuracy for the M1 metric and an \textbf{87.57\%} for the M3 metric in provisional diagnosis. This contrasts with average physician accuracies of \textbf{50.27\%} for M1 and \textbf{62.27\%} for M3 (\textit{Figure} \ref{fig:m1_m3}). Similarly, for specialization recommendations, AmarDoctor achieved \textbf{91.35\%} accuracy in the M1 metric, outperforming the average physician accuracy of \textbf{62.6\%} (\textit{Figure} \ref{fig:s1}). These results highlight the system's capacity to provide robust preliminary diagnostic and referral guidance, which is invaluable in resource-limited settings, and align with similar studies where AI-based diagnostic models have demonstrated high accuracy compared to human physicians \textcolor{blue}{\cite{Schmieding2022}, \cite{Norgeot2019}, \cite{gilbert2020accurate}, \cite{Peven2023}, \cite{Zeltzer2023}}.

The ability of such a virtual consultation tool to potentially deliver healthcare to millions currently deprived of adequate medical access represents a significant advancement.

Furthermore, our analysis of diagnostic discrepancies revealed nuanced insights into the system's performance. The \textbf{M3} metric identified \textbf{23} failure cases in provisional diagnoses out of \textbf{185} cases. Interestingly, \textbf{20} of these instances showed concordance between AmarDoctor's diagnoses and those of the physicians, despite diverging from the gold standard. This shared discrepancy, exemplified by cases such as \textbf{Patient 37} (Respiratory infection vs. Asthma) and \textbf{Patient 253} (Renal disorder due to type 1 diabetes mellitus vs. Type 1 diabetes mellitus), underscores the inherent complexities of differential diagnosis based solely on symptom data. Such cases often involve considerable symptom overlap or conditions where primary and secondary manifestations are intertwined, requiring deeper clinical context for precise differentiation. The alignment of AmarDoctor's recommendations with a consensus of three or more physicians in \textbf{10} of these cases (\textit{Table} \ref{tab:vignette1}) suggests that the AI system is learning and reflecting patterns of clinical reasoning, even in challenging scenarios where human experts also face ambiguity. This reinforces its role as a supportive tool that can offer valuable data-driven insights to physicians, potentially highlighting areas for further investigation.

While our approach demonstrates promising results and significant potential, several limitations warrant consideration and provide avenues for future work:

\subsection{Language Limitations} 
The current language coverage restrict the platform's initial version to Bengali and English-speaking populations. While our system successfully mapped \textbf{908} unique symptoms to over \textbf{3,237} Bengali (both standard and colloquial) variations across major dialects like Sylheti and Chittagonian (as detailed in Section 2.6.2), its broader impact within the South Asian region would be amplified by adaptation to other widely used languages. This expansion would require extensive linguistic and medical vocabulary curation for each new language, building upon our established methodology.

Another technical limitation is specific to the voice-interactive assistant, \textbf{'Aisha'}, which functions as the app's main gateway. To understand user intent for general navigation—such as finding a hospital or downloading prescriptions—this gateway chatbot currently relies on Google's standard Speech-to-Text model. We have observed that this model is somewhat limited in accurately capturing the phonetic nuances of diverse Bengali dialects, which can lead to transcription errors that hinder the intent classifier's performance.

In contrast, our AI-guided symptom reporting module leverages our own proprietary Bengali medical ASR model to ensure higher accuracy for clinical data capture. Future work will focus on replacing the standard STT in the gateway chatbot with our proprietary ASR technology, thereby creating a consistently accurate and reliable voice experience across all platform features.

\subsection{Data Limitations} 
Data limitations are inherent in the development of any AI-driven healthcare system. The effectiveness and generalizability of AmarDoctor are directly dependent on the quality and comprehensiveness of its underlying health knowledge graph, which was constructed from 1.4 million patient records (Section 2.1). Enriching this knowledge base to include a wider range of diseases, symptoms, and treatment options, particularly those prevalent in diverse South Asian populations, is crucial. Such enrichment would directly improve the diagnostic accuracy metrics (M1 and M3) observed in our results, especially for less common conditions or those with subtle presentations. 

Another notable limitation is the lack of comprehensive patient data integration in the provisional diagnosis generation system. The provisional diagnosis generation process does not yet fully incorporate patient demographics (e.g., age, gender beyond basic distribution), medical history, or lifestyle factors. Integrating this broader clinical context is crucial for generating more comprehensive, accurate, and personalized diagnostic suggestions, as these factors profoundly influence disease presentation and risk. For instance, deeper integration of patient history might have further refined diagnoses in cases like \textbf{Patient 253}, where the distinction between primary and secondary conditions was challenging even for physicians. Future iterations will aim to seamlessly integrate patient demographics (e.g. age, gender), medical history, and lifestyle factors into the provisional diagnosis generation process to enhance diagnostic precision and generate more comprehensive, accurate, and personalized suggestions.

\subsection{Complexity of Disease Manifestation} 
A significant challenge arises from diseases that present with very similar and overlapping symptoms, as shown by some of the discrepant cases in our results. Medical conditions characterized by intricate or atypical symptom patterns, like the misclassification of Respiratory infection as Asthma for \textbf{Patient 37} or the overlap between Renal disorder and Type 1 diabetes for \textbf{Patient 253}, may not be easily captured or fully elucidated through a linear question-and-answer format. To effectively handle such complex symptom interactions and improve the system's diagnostic granularity, future research should explore more advanced AI techniques, such as sophisticated natural language processing (NLP) models capable of understanding nuanced patient narratives and advanced machine learning algorithms designed for complex pattern recognition in medical data. This would allow the system to move beyond a structured query approach to interpret more free-form symptom descriptions, potentially resolving ambiguities observed in our evaluation.

\subsection{Limited Scope of Clinical Evaluation}

The \textbf{number of physicians involved} in both the curation of patient vignettes and the evaluation of the platform's accuracy presents a scope limitation. While the clinical consultants (\textbf{STK} and \textbf{SD}) are highly experienced (with \textbf{7} and \textbf{12} years of experience respectively, as mentioned in Section 2.3), a larger, more diverse pool of physicians, ideally with at least six years of clinical experience, would enhance the robustness and generalizability of the curated vignette cases. Similarly, assessing the platform's accuracy against a larger number of more experienced physicians (beyond the five with an average of three years of experience, as noted in Section 2.4) would further strengthen the system's reliability and applicability across a wider clinical spectrum, providing even more compelling evidence for the reported M1 and M3 accuracies. In the next iteration we will conduct a systematic evaluation of AmarDoctor's provisional diagnosis system against doctor's diagnosis on a larger pool of consented anonymized real patient data, who received virtual consultation on AmarDoctor platform.

\subsection{Challenges of Deploying Digital Health Systems}

Finally, the deployment of AI in healthcare requires careful consideration of \textbf{ethical implications}. Issues related to data privacy, algorithmic bias, and accountability are paramount. While AmarDoctor prioritizes safety and confidentiality in data processing, continuous development of robust ethical frameworks and guidelines is essential to ensure the responsible, equitable, and trustworthy deployment of such powerful diagnostic tools, particularly when addressing diverse and vulnerable populations. This includes ongoing monitoring for biases that might arise from data imbalances or algorithmic design, and ensuring transparency in AI-driven recommendations, especially in cases of discrepancy as observed in our results.

\subsection{Future Improvements}
To further enhance the functionality of the platform and address the limitations mentioned above, our future work will focus on several key directions. First, we will enrich our predictive model by incorporating a wider range of patient data, including medical histories, medications, and diagnostic test results. To promote wider adoption, we are also developing easily integratable plugins to connect the AmarDoctor system to popular electronic health record (EHR) platforms, ensuring that our AI recommendations can function seamlessly across the broader healthcare spectrum. Second, in collaboration with local and global healthcare partners, we are initiating large-scale randomized controlled trials in various low-resource settings. These studies are crucial for comprehensively assessing the system's real-world effectiveness and generalizability and for identifying areas for iterative improvement in practical clinical environments. Third, we plan to develop specialized modules for specific, complex disease areas such as cardiology and oncology. This will tailor the system to the nuanced needs of different patient populations and improve diagnostic accuracy within these domains. Finally, we are continuously investigating emerging technologies, including advanced machine learning and deep learning architectures, to build capabilities for early diagnosis and personalized treatment recommendations. A primary focus will be on adapting these sophisticated models for diverse populations and integrating them seamlessly into the AmarDoctor platform for delivery to both patients and providers.

By systematically following these research and development directions, the AmarDoctor platform can be refined to become an even more impactful tool to improve healthcare access, efficiency, and patient outcomes globally.

\section{CONCLUSION}
In this study, we introduced \textbf{AmarDoctor}, a groundbreaking multilingual digital healthcare platform designed to enhance primary care access and patient management for Bengali speakers. This platform uniquely integrates AI-driven symptom assessment with clinical decision support, serving as a crucial machine-in-the-loop tool. Our comprehensive evaluation, utilizing a standardized set of \textbf{185} clinical vignettes and comparing system outputs against practicing physicians, demonstrated AmarDoctor's robust capabilities in generating provisional diagnoses and appropriate specialization recommendations. The patient-centric design, featuring multimodal input and support for diverse Bengali dialects, significantly addresses challenges related to digital literacy and linguistic barriers prevalent in underserved communities. The platform's pilot launch has already facilitated thousands of virtual consultations, with user feedback highlighting its profound positive influence on healthcare accessibility and its potential to democratize healthcare in low-resource settings. By streamlining the triaging process and providing structured clinical insights, AmarDoctor offers a scalable solution to optimize healthcare delivery, reduce consultation burdens, and improve patient outcomes, particularly in low-resource settings. This work establishes a foundational step towards more inclusive and accessible AI-driven healthcare solutions globally, with a clear vision for continued expansion and refinement based on real-world impact.
To amplify its broader impact within the South Asian region, future work will involve adaptation to other widely used languages, requiring extensive linguistic and medical vocabulary curation for each new language, building upon our established methodology.

\section*{ACKNOWLEDGEMENTS}
We extend our gratitude to Shadman Shaqif (Junior Software Engineer, MedAi Limited) and Md. Maruf Hasan (App Developer, MedAi Limited) for their technical contributions in developing the AmarDoctor web platform and mobile application, respectively. We are also deeply grateful to \textbf{MIT Solve}, \textbf{The Roddenberry Foundation}, and \textbf{Google for Startups Sustainable Development} program for their invaluable support and funding.

We would also like to acknowledge the following physicians for their contributions during the development and evaluation of the AmarDoctor platform:

\textbf{Dr. Suparna Das} (for reviewing the vignettes)\\
\textbf{Dr. Moushumi Paul} (for reviewing the vignettes)\\
Dr. Zakia Farhana Rahman\\
Dr. Meem Islam\\
Dr. Md. Intisher Akhter Talukdar\\
Dr. Mahzabeen Hossain Bidushi, and\\
Dr. Rubaiyat Karim Mou


\bibliographystyle{unsrt}  
\bibliography{AmarDoctor}

\end{document}